\documentclass[conference,10pt,twocolumn]{IEEEtran} 

\usepackage{cite} 
\usepackage{graphics}
\usepackage{graphicx}
\usepackage[english]{babel}
\usepackage{amsmath}
\usepackage{amsthm}
\usepackage[table,xcdraw]{xcolor}
\usepackage[linesnumbered, ruled]{algorithm2e}
\usepackage[noend]{algpseudocode}
\usepackage[font=footnotesize]{caption}
\usepackage{epsfig}
\usepackage{xfrac}
\usepackage{nicefrac}
\usepackage{psfrag}
\usepackage{times}
\usepackage{balance} 
\usepackage{amssymb}
\usepackage{amsfonts}
\usepackage{bm}
\usepackage[bottom]{footmisc}
\usepackage{mwe}
\usepackage{color}
\usepackage{booktabs,multirow}
\usepackage{tabularx}
\usepackage{tikz}
\usepackage{kotex}
\usetikzlibrary{patterns,arrows}
\usepackage{physics}
\usepackage{verbatim}
\usepackage{subcaption}

\newcommand{\tredd}[1]{\textcolor{black}{#1}}

\SetKwRepeat{Do}{do}{while}
\SetKwInOut{Input}{Input}
\SetKwInOut{Output}{Output}

\definecolor{color1}{RGB}{228,26,28}
\definecolor{color2}{RGB}{55,126,184}
\definecolor{color3}{RGB}{77,175,74}
\definecolor{color4}{RGB}{152,78,163}
\definecolor{color5}{RGB}{255,127,0}
\definecolor{color6}{RGB}{200,200,200}

\begin{document}

\title{\fontsize{19}{24}\selectfont PMNet: Robust Pathloss Map Prediction via Supervised Learning}

\author{ 
Ju-Hyung Lee, Omer Gokalp Serbetci, Dheeraj Panneer Selvam, and Andreas F. Molisch \\

	\small Ming Hsieh Department of Electrical and Computer Engineering, University of Southern California, 
	\small Los Angeles, USA \\
        \small \{juhyung.lee, serbetci, dpanneer, molisch\}@usc.edu
	}

\maketitle



\begin{abstract}
Pathloss prediction is an essential component of wireless network planning. While ray tracing based methods have been successfully used for many years, they require significant computational effort that may become prohibitive with the increased network densification and/or use of higher frequencies in 5G/B5G (beyond 5G) systems. 
In this paper, we propose and evaluate a data-driven and model-free pathloss prediction method, dubbed \emph{PMNet}.
This method uses a supervised learning approach: training a neural network (NN) with a limited amount of ray tracing (or channel measurement) data and map data and then predicting the pathloss over location with no ray tracing data with a high level of accuracy.
Our proposed pathloss map prediction-oriented NN architecture, which is empowered by state-of-the-art computer vision techniques, outperforms other architectures that have been previously proposed (\textit{e.g.}, UNet, RadioUNet) in terms of accuracy while showing generalization capability.
\tredd{
Moreover, PMNet trained on a 4-fold smaller dataset surpasses the other baselines (trained on a 4-fold larger dataset), corroborating the potential of PMNet.\footnote{
The trained model and codes are publicly available on the Github page: {https://github.com/abman23/PMNet}
}
}
\end{abstract}

\section{Introduction} 



In order to provide the higher data rates, greater reliability, and shorter latency required by new applications, 5G and Beyond 5G (B5G) systems use a combination of higher deployment density, distributed architectures such as cell-free massive MIMO (CF-mMIMO), and transmission in high frequency bands. At the high frequencies, the details of the environmental structures including buildings, vegetation, roads, and other myriads of urban terrain features, become much more important. At the same time the large increase in infrastructure nodes due to network densification and CF-mMIMO requires dramatically faster methods for cell planning/optimization, such as pathloss prediction - the traditional ray launchers are too slow for repeated runs inherent in the optimization process. Thus, there is an unprecedented need for accurate and fast pathloss prediction over large-scale terrain map data.


\subsection{Related Works: Pathloss Map Prediction}

Pathloss map prediction (PMP) estimates a certain location's propagation path loss through a model with or without a measurement data. Since actual field measurements with channel sounders are too expensive and complicated on a large scale, ''measurements" have long been replaced by computer simulations such as ray tracing \cite{kim1999radio} and ray launching \cite{WirelessInsite}.\footnote{In the remainder of this paper we will use ''measurements", "ray tracing", and "ray launching" equivalently, indicating a suitable method for finding a location-specific ground truth for the pathloss.} Over the past 30 years, both the efficiency and accuracy of ray tracing\footnote{In line with much of the literature, we henceforth use here the word "ray tracing" even for algorithms that are strictly speaking "ray launching"} have increased remarkably \cite{degli'Esposti2007}, and computations are helped by the prevalence of GPUs (graphic processing units) that are well suited for ray tracing tasks.

Still, due to the factors mentioned above (higher frequencies requiring more detailed environmental data bases; higher deployment density require larger-scale simulations), ray tracing is too computationally intensive for large-scale deployment of B5G. Therefore, simplified models like the dominant path model \cite{dominant}, or fine-tuning of generic pathloss models with limited measurement data \cite{voronoi, survey_radiomap} have been proposed over the years, though they have found only limited acceptance by network operators for large-scale planning.

In recent years, supervised machine learning (ML) has been successfully used for a large number of challenging classification and regression problems in a variety of areas, ranging from natural language processing to data packet scheduling. This has motivated the investigation of ML for pathloss prediction and deployment planning.  Trained by a map of the environment (topology/morphology) and a relatively small set of measurement data, it learns how to do pathloss prediction when only an environmental map is available. 
In particular, according to \cite{ref:RadioUNet} and \cite{ref:Fadenet}, pathloss prediction is formulated as a pixel-wise linear regression problem. The regression vision task is performed individually for each pixel.
\tredd{
The authors in the aforementioned works leverage the convolutional neural network architecture with computer-vision techniques and use topographic information (\textit{e.g.}, terrain, building, and foliage heights) as their input feature for PMP.
In addition, several studies have been followed, such as implanting conventional ML architectures (\textit{e.g.}, GAN) \cite{rmp_gan}, exploiting real-measurement data in the training dataset \cite{rmp_manhattan}, or focusing on a feature-engineering for the dataset of PMP instead of the PMP task itself \cite{image_texture}.
Our view is aligned with the former angle, which will be elaborated in Sec. \ref{section:background}. 
}

\subsection{Major Contribution}


Our major contributions are summarized as follows.
\begin{itemize}
\item 
\tredd{
We design a PMP-oriented neural network architecture (dubbed PMNet) by utilizing computer-vision techniques, which achieves up to 3.01x higher prediction accuracy over other methods;} this evaluation is based on on our ray tracing simulation data on the USC campus map (see \textbf{Table \ref{Table_Comparison_Accuracy}} in Sec. \ref{Numerical Result}).

\item 
\tredd{
Considering the scarcity of channel measurement data, our proposed PMNet is validated with small dataset; besides, other ray tracing simulation datasets (\textit{RadioMapSeer} \cite{ref:RadioUNet}) is also used to show generalizability of the PMNet, \textit{i.e.}, if this network is working well with other datasets (see \textbf{Table \ref{Table_Comparison_Small}} and \textbf{\ref{Table_Comparison_RadioMapseer}} in Sec. \ref{Numerical Result}). 
Moreover, we present an example of the application of such high-accuracy achieving PMNet (see \textbf{Fig. \ref{fig:Coverage Map}} in Sec. \ref{Numerical Result}).
}

\end{itemize}


\section{Background} \label{section:background}

\subsection{CNN and UNet}

In the realm of supervised learning, convolutional neural networks (CNNs) are the go-to networks for image classification, due to their lower computational complexity with fewer nodes compared to fully-connected networks.

A standard CNN consists of convolution layer, non-linear activation function, pooling, and a fully-connected layer. 
The convolutional layer uses filters or kernels $h_{i, j}$ to apply 2-D convolutional operations to the input feature $\boldsymbol{x} \in \mathbb{R}^{m \times n}$ as shown below: 
\begin{equation}
    z_{i,j} = \sum_{m = 1}\sum_{n = 1} h_{i - m, j - n}\vb*{x}_{m, n}
\end{equation}
After passing the convolutional layer, the input features shrink to smaller feature maps while sharing the parameters along the convolution procedure (\textit{e.g.}, sparse interactions \cite{deep_book}), which leads to efficient computation for multi-dimensional data.
CNNs work as the widely used backbone architecture for image classification,\textit{e.g.}, AlexNet \cite{alexnet} and ResNet \cite{resnet}. 

UNet bases its architecture upon a combination of CNNs with autoencoder (AE) architecture \cite{auteoncoders}. AEs are neural networks that aim to find a latent space that represents the data in a smaller space; thus, AEs are widely used for dimensionality reduction \cite{dim_redux} and generative modeling \cite{vae}. UNet uses the autoencoder architecture for its encoder and decoder parts. 
A defining feature of UNet is the \emph{U}-shape connection between this encoder-decoder.
Traditionally CNNs are used to classify an image as a whole, but UNet makes pixel-wise classification to get the desired semantic segmentation.

\subsection{RadioUNet and FadeNet}

RadioUNet \cite{ref:RadioUNet} and FadeNet \cite{ref:Fadenet} are two main NNs that have been used for ML-based PMP; both employ U-Net as backbone for PMP. 
Each uses the U-Net architecture differently; FadeNet uses a single UNet with $28$ convolutional layers with some slight modification, while RadioUNet uses two cascaded U-Nets with $18$ convolutional layers each called W-Net.

While FadeNet uses a general NN training process, RadioUnet employs curriculum learning where each UNet is trained separately, requiring more computation but achieving better training results.
Both networks consider the normalized mean square error(NMSE), and the root mean square error(RMSE) for their performance metric. 
Main input features are \textit{i)} city map and \textit{ii)} TX locations, with additional features (\textit{e.g.}, cars and foliage heights, and line of sight (LoS) condition).



\section{Pathloss Map Prediction} \label{Body}

\begin{figure}[h!]
  \centering
  \begin{subfigure}[t]{.24\textwidth}
    \centering
    \includegraphics[width=1\linewidth]{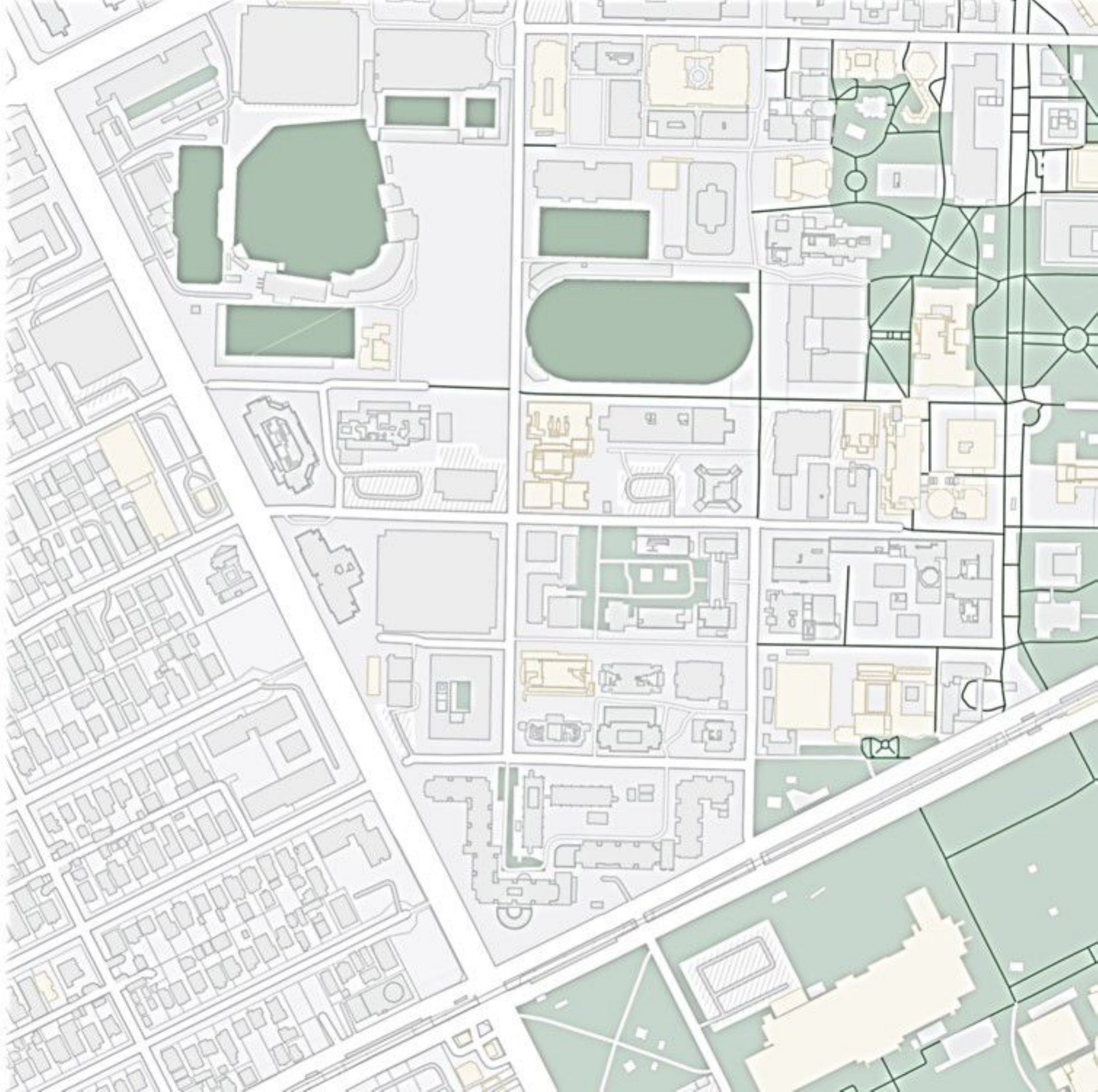}
    \caption{University of Southern California (USC) campus map.}
    \label{fig:USC_Map_Morphological}
  \end{subfigure}\hspace*{.02\textwidth}%
  \begin{subfigure}[t]{.24\textwidth}
    \centering
    \includegraphics[width=1\linewidth]{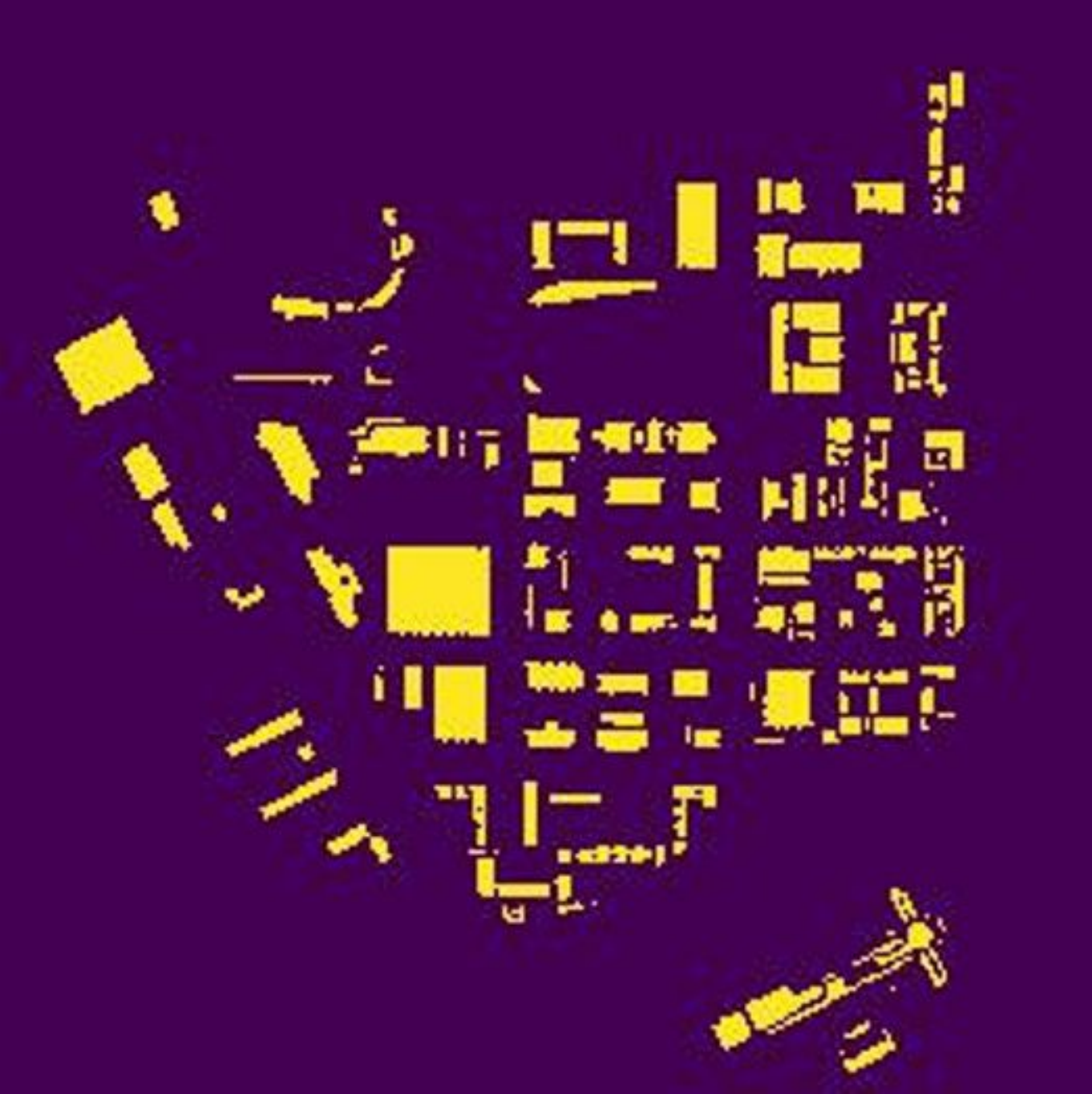}
    \caption{Building map of USC}
    \label{fig:USC_Map_Building}
  \end{subfigure}
  \caption{University of Southern California (USC) campus map. Fig. \ref{fig:USC_Map_Morphological} (Left) is imported and converted to Fig. \ref{fig:USC_Map_Building} (Right). With that, the ground-truth pathloss map over the USC campus is obtained by using the \emph{Wireless Insite} ray tracing simulation and pre-processed using interpolation, grey conversion, and data augmentation methods.}
  \vspace{-1.0em}
  \label{fig:USC_Map}
\end{figure}

In this section, we propose a pathloss map prediction-oriented NN architecture, dubbed PMNet.
The processes of dataset preparation, design of PMNet, and evaluation are elaborated in the following subsections.

\subsection{Dataset}
The ground-truth (GT) dataset is obtained by the commercial ray tracing tool \emph{Wireless Insite} \cite{WirelessInsite}, taking into account the geographical and morphological features of propagation environment. Then, the dataset is prepared after pre-processing (\textit{e.g.}, interpolation and data-augmentation methods).

\subsubsection{ray tracing simulation}
In ray tracing simulations, rays from specified TX antenna gradually lose power over distance, due to free space power loss.  
Further attenuation occurs when those rays intersect with vegetation, and during reflections from, or transmission into, buildings. 

Following such physical principles, ray tracing emulates the behavior of each multipath component (MPC) between TX and RX, including the free-space power loss and interaction with different interacting objects (IOs). This allows to compute for each MPC the information of absolute amplitude $|a|$, phase $ \phi$, directions of departure $\Omega$ and arrival $\Psi$, and the delay $\tau$. 
The contribution of $m$-th MPC can be expressed as:
\begin{align}
    h_{m}(\tau, \Omega, \Psi) = |a_{m}| e^{j \phi_{m}} \delta(\tau - \tau_{m}) \delta(\Omega - \Omega_{m}) \delta(\Psi - \Psi_{m}).
\end{align}
The sum of contributions from all MPCs is given by 
\begin{align}
    h(\tau, \Omega, \Psi) =  \sum_{m=1}^{N} h_{m}(\tau, \Omega, \Psi).
\end{align}
The path gain (inverse of the path loss) averaged over the small-scale fading can be computed as
\begin{align}
    \mathrm{PG} =  \sum_{m=1}^{N} | h_{m}(\tau, \Omega, \Psi)|^2 .
\end{align}


Our pathloss map uses the information of path gain (represented on a dB scale) while other information on angles and delay are not (though this information can be used for the further applications involving, \textit{e.g.}, beamforming algorithms). 

For our ray tracing simulation, the actual geographical and morphological map of the University of Southern California is considered  (see Fig. \ref{fig:USC_Map} and Table \ref{table_Paramter} for more details). While the simulations are performed at 2.5 GHz to emulate the most widely used cellular band, similar simulations can be performed in other frequency bands as well with minor parameter adjustments.

  

\subsubsection{Pre-processing}
The raw dataset from the ray tracing simulation is pre-processed using \emph{interpolation} and \emph{data augmentation} methods, and is then divided into training, validation, and testing datasets.

\textbf{Interpolation}.\quad
The ray tracing simulations are carried out over a discrete set of Rx locations, and it is computationally challenging to perform these simulations for every location (pixel). Consequently, some pixels may lack channel information. To interpolate the missing data for map pixels, this study utilizes \emph{bilinear interpolation}, which effectively approximates values between adjacent locations while preserving computational efficiency.


\textbf{Grey conversion}.\quad
The map with the channel information, which is our GT data, is converted to a binary grey-scale $256\times256$ size image.
\tredd{
For each pixel, the received power $\vb*P_{\mathrm{RX}}$ (in [dBm]) is converted to values between $0$ and $1$ using Min-Max normalization. The minimum value was chosen as -150 dB, since smaller values of the pathgain are irrelevant for system considerations. }
Note that we set the pixels at which buildings are present as $0$, since we do not consider indoor coverage. 
Then, our GT map becomes one channel $256\times256$ image. 
Again, the pixels located for the building are filled with $0$, and our region-of-interest (RoI) is filled with values between $0$ and $1$, which corresponds to $\vb*P_{\mathrm{RX}}$.

\textbf{Data augmentation}.\quad
Generally,  a larger data set is required in order to achieve a better performance of NN training. 
In other words, the larger the data set, the better. 
Here, we use mainly two augmentation methods: cropping and rotation. 
The entire map data is cropped into images of about a quarter of the size taking TX as an anchor point, augmenting the size of the dataset by a factor of $96$. 
The image is first cropped as $64\times64$ size image and then upsampled to $256\times256$ size image. 
Note that some cropped images, not including any TX, are skipped since the TX location will be used as our second input channel.
After cropping, the image sets are rotated by $90^{\circ}$, $180^{\circ}$, and $270^{\circ}$, thus increasing the size of the dataset by a further factor 4.

\textbf{Sampling}.\quad 
We consider two types of datasets: Exclusive and Random.
The size of each dataset is the same, namely $19,016$ images, but the way of sampling is different. 
We adopt an Exclusive sampling scheme for dividing the dataset into training and validation sets. Specifically, 90\% of the images are randomly assigned to the training set, while the remaining 10\% constitute the validation set, ensuring that images from the same map are exclusively included in one set. 
In contrast, the Random and Random (Small) sampling methods randomly allocate images to the training and validation sets, allowing for the possibility that images from the same map (but with distinct TX antenna locations) may be present in both sets. Although this approach offers a broader range of samples, it may also lead to a higher degree of correlation between the training and validation sets.

\begin{table}[]   
  \centering
  \resizebox{1\columnwidth}{!}{\begin{minipage}[t]{0.5\textwidth}
  \caption{Parameters of network environment in USC dataset.}
  \centering
  \begin{tabular} {l l}
	\toprule[1pt]
	\textbf{Parameter} & \textbf{Value} \\
	\cmidrule(lr){1-1} \cmidrule(lr){2-2}
	USC campus map scale & $920 \times 920$ {[}m$^2${]} \\  
	Map scale of RoI (per pixel) & $243 \times 243$ {[}m$^2${]} ($0.95 \times 0.95$ {[}m$^2${]}) \\  
	Carrier frequency & $2.5$ {[}GHz{]} \\
	Bandwidth & $1$ {[}MHz{]} \\ 
	Transmit power & $30$ {[}dBm{]} \\
	TX antenna type & Isotropic (vertical) \\ 
	TX waveform & Sinusoid \\ 
    \midrule[.7pt]
	\textbf{Dataset} & \textbf{\# of Training (Validation) Data} \\
	\cmidrule(lr){1-1} \cmidrule(lr){2-2} 
	Random  & $1.52 \times 10^{4}$ $(3.8 \times 10^{3})$ \\ 
	Exclusive & $1.52 \times 10^{4}$ $(3.8 \times 10^{3})$ \\ 
	Random (Small) & $0.35 \times 10^{4}$ $(0.75 \times 10^{3})$ \\ 
	\bottomrule[1pt]
\end{tabular}

  \label{table_Paramter}
  \end{minipage}}
  \vspace{-1.0em}
\end{table}

\subsection{Design of Network Architecture}

We here introduce the design process of our proposed PMNet.
Our design principles are summarized as follows: 
\textit{i)} several state-of-the-art techniques in the field of image processing are carefully selected and tested, \textit{ii)} some essential techniques are selected following the concept of ablation study, and \textit{iii)} the NN with selected techniques is optimized with many trials.
The considered essential techniques are elaborated in the following.

\textbf{Atrous convolution}.\quad
The \emph{receptive field}, the size of the region of the input that produces the feature, corresponds to the resolution of features computed by convolutional layer and the field-of-view (FoV) of the filter.
It is known that there is a logarithmic relationship between the localization accuracy and the use of context (receptive field size), that suggests that large receptive fields are necessary for wide-level recognition tasks.
In other words, the size of the receptive field should be sufficient if the given dataset and task should be observed with wide FoV.

A standard convolutional filter detects a particular feature by sliding over the input feature map, resulting in the output feature map seeing only the adjacent part of the input feature map.
In terms of computational complexity, it is expensive to have a wide receptive field with the standard convolutional filter. Thus, broadly speaking, the receptive field of the standard convolution filter is somewhat narrow thus seeing only little context.

\emph{Atrous} (or \emph{dilated}) convolution \cite{ref:Deeplabv3+} allows to capture a larger context with a wider FoV by modifying the standard convolution operation. 
For the two-dimensional case, for each 2D location $i$-$j$, atrous convolution is applied over the input feature map $f$ on the output feature map $g$ with the convolution filter $w$, which can be expressed as follows:
\begin{equation}
    g_{i, j} = \sum_{m=1}^{k}\sum_{n=1}^{k} f_{i+ rm, j + rn}  w_{m, n}, \label{eq:AstrousConvolution}
\end{equation}
where the atrous rate $r$ determines the stride and $k$ represents the kernel size. 
One can understand that the standard convolution is a special case of \eqref{eq:AstrousConvolution} where $r = 1$. 
Note that the FoV of filter can be adaptively controlled by $r$.

\textbf{Hourglass networks}.\quad
The convolutional autoencoder network is a widely applied architecture for many computer vision tasks, \textit{e.g.}, object detection, human pose estimation, and semantic segmentation \cite{ref:Deeplabv3+}, to learn a lower-dimensional representation from a higher-dimensional dataset. 
Generally, an encoder module gradually reduces the feature maps and captures higher semantic information while a decoder module gradually recovers the spatial information. 
However, as the encoder shrinks the input feature, a bottleneck problem occurs where not all features can be encoded.
Several architecture, including U-Net \cite{ref:UNet}, build upon additional \emph{connections} between the encoder and the decoder parts (called hourglass architecture) to overcome the bottleneck problem.
By doing this, the upsampling part contains a larger number of feature channels, which allow the network to propagate context information to higher resolution layers.

\begin{figure}[t!]
    \centering
    \includegraphics[width=1\columnwidth]{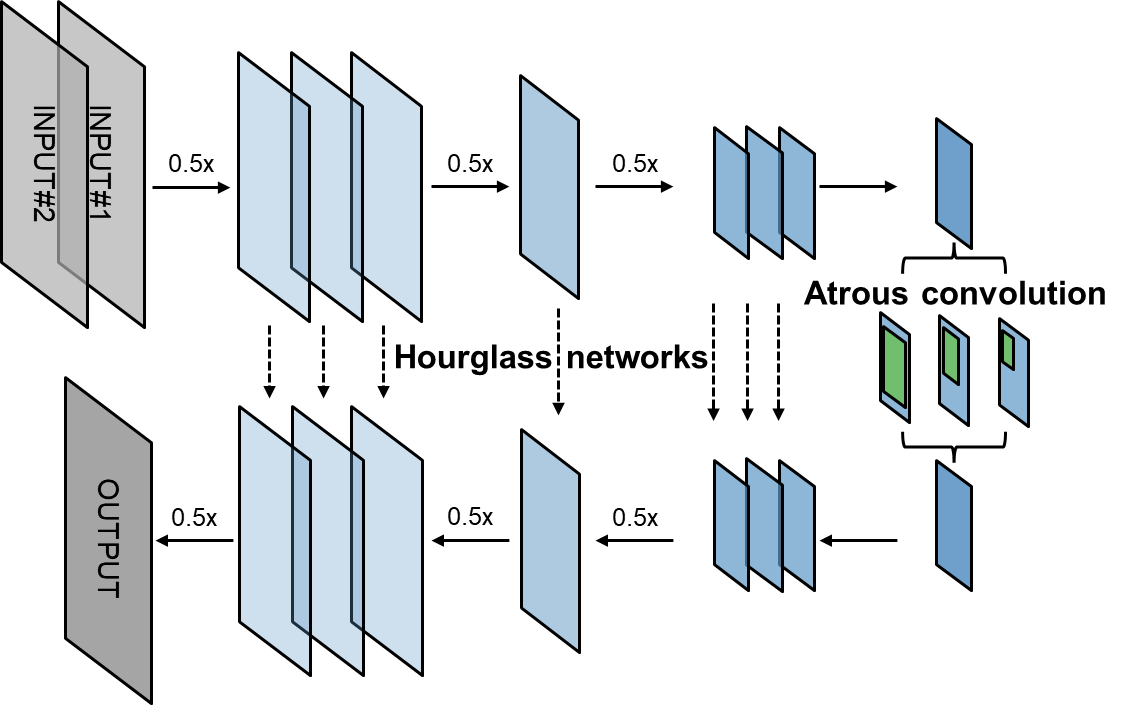}
    \caption{Illustration of our proposed PMNet architecture.}
    \label{fig:NN architecture}
    \vspace{-1.0em}
\end{figure}

\textbf{Design choices}.\quad
For our pathloss map prediction, the network needs to do image segmentation. 
That is, it needs to find the RoI, \textit{i.e.}, the exterior part of a building; then, it needs to predict the received power over the RoI, considering the reflection/scattering behaviors from different IOs in a given propagation environment.

Our proposed PMNet is designed based on such methods, atrous convolution and hourglass network.
The atrous convolution allows to probe convolutional features at multiple scales, which facilitates the network to capture a broader context in the map data.
Besides, the hourglass network is shaped more or less symmetric, which enables to efficiently propagate context information from the encoder side to the decoder side \cite{ref:UNet}.
Building on top of those ideas, the PMNet precisely predicts the pathloss map even with a limited data set; see Fig. \ref{fig:NN architecture} for the details of the PMNet architecture.

\subsection{Evaluation}

\textbf{Normalized mean square error (NMSE)}.\quad
The NMSE emphasizes the level of misestimation in the entire dataset. 
Lower values of NMSE denote better performance and vice versa. The expression of NMSE is given below over images $\vb* x$ and $\vb* y$:            
\begin{align}
    \mathrm{MSE}(\vb* x,\vb* y) =& \frac{1}{N} \sum_{m}(\vb* x_{m} - \vb* y_{m})^{2} \\
    \mathrm{NMSE}(\vb* x,\vb* y) =& \frac{\mathrm{MSE}(\vb* x,\vb* y)}{\mathrm{MSE}(\vb* x,0)} = \frac{||\vb* x-\vb* y||^{2}_{2}}{||\vb* x||^{2}_{2}}.
\end{align}
For our PMP task, each pixel value, representing a power value at that particular coordinate, is a principal indicator for accuracy. 
Thus, NMSE works considers the normalized value of every pixel for the computation of the accuracy. 

\textbf{Normalized depth loss (NDL)}.\quad
Normalized depth loss (NDL) is a good measure used to compare variations in depth structures in images as proposed in \cite{ref:gradientloss, ref:Multiloss}. 
NDL calculates the logarithm of depth errors, which is sensitive to shifts in depth direction. 
Let $\mathrm{err}_m$ $\triangleq$ $||\vb* x_{m} - \vb* y_{m}||_1$ . 
The loss is calculated as follows :
\begin{align}
    \mathrm{NDL} =& 
    \dfrac{1}{N} \sum^{N}_{m=1} F(\mathrm{err}_m)
\end{align}
where, $F(\vb* x) = \ln(\vb* x +  \vb*\gamma), \vb*\gamma \geq  0$.
Just as for the NMSE, each pixel's value is of concern rather than the entire image. 
\tredd{
It is an effective metric for seeing its performance in detecting if a certain pixel is RoI or not, thanks to its depth direction sensitivity \cite{ref:gradientloss}.
}



\section{Numerical Results} \label{Numerical Result}

\begin{table*}[!h]
\centering
\caption{Comparison of UNet, RadioUNet, and our proposed PMNet in terms of NMSE and NDL.}
\resizebox{1.5\columnwidth}{!}{\begin{minipage}[h]{1.33\columnwidth}
\centering
\label{Table_Comparison_Accuracy}
\begin{tabularx}{1\linewidth}{l c c c c}
\toprule[1pt]
Dataset & \multicolumn{2}{c}{ Random} & \multicolumn{2}{c}{Exclusive}  \\
\cmidrule(lr){1-1} \cmidrule(lr){2-3} \cmidrule(lr){4-5}
{Scheme} & Average $\mathrm{NMSE}$ & Average $\mathrm{NDL}$ &  Average $\mathrm{NMSE}$ & Average $\mathrm{NDL}$ \\
\cmidrule(lr){1-1} \cmidrule(lr){2-2} \cmidrule(lr){3-3} \cmidrule(lr){4-4} \cmidrule(lr){5-5}

UNet \cite{ref:UNet} & $0.0042280$ \; & $0.0399405$ & $0.0082982$ \; & $0.0649527$ 
\\
RadioUNet \cite{ref:RadioUNet} & $0.0009335$ \; & $0.0185123$ & $0.0046352$ \; & $\textbf{0.0451356}$ 
 \\
PMNet (Proposed) & $\textbf{0.0003099}$ \; & $\textbf{0.0111533}$ & $\textbf{0.0045575}$ \; & $0.0453891$ 
 \\

\bottomrule[1pt]
\end{tabularx}

\vspace{-1.em}
\end{minipage}}
\end{table*}

\begin{figure*}[!h]
  \centering
  \begin{subfigure}[t]{.255\textwidth}
    \centering
    \includegraphics[width=\linewidth]{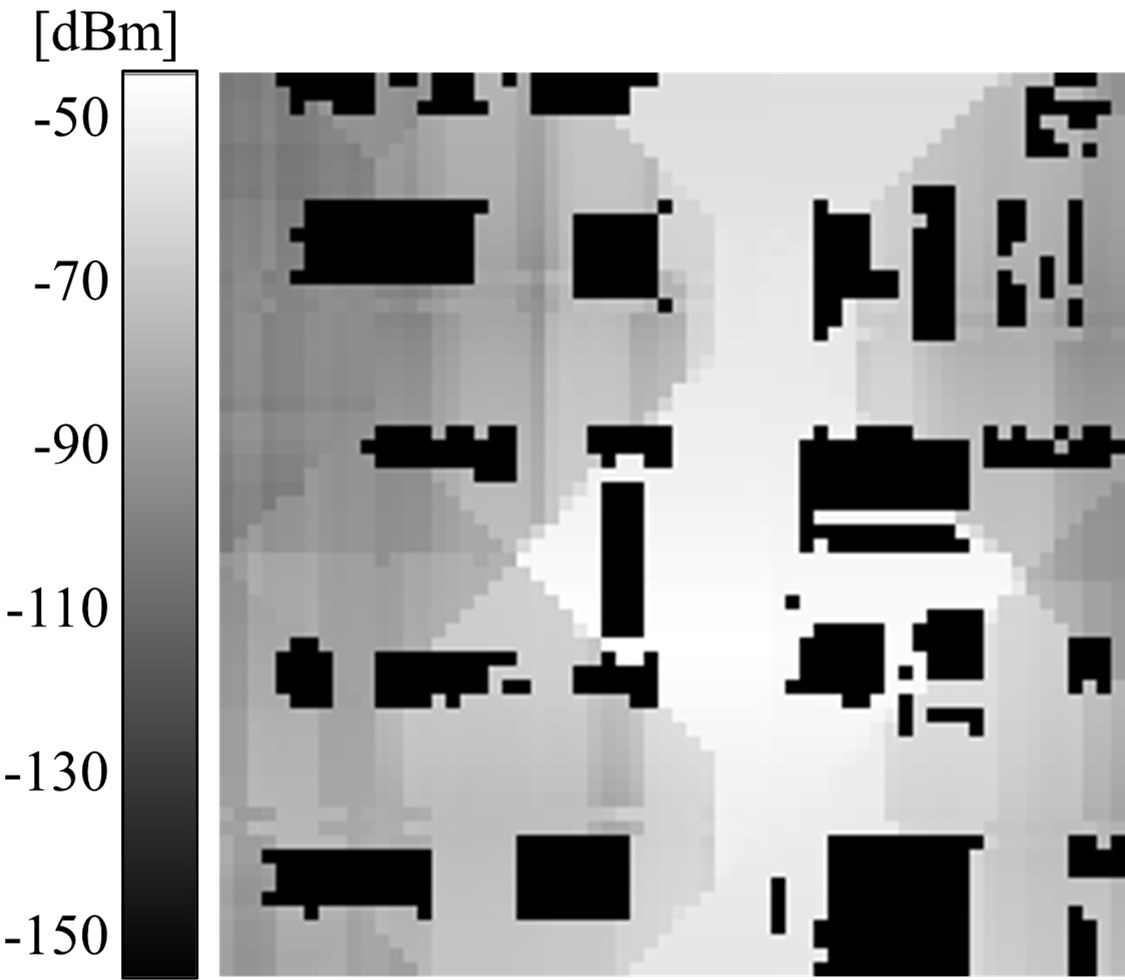}
    \caption{Ground Truth}
    \label{Comparion:GT}
  \end{subfigure}\hspace*{.025\textwidth}%
  \begin{subfigure}[t]{.205\textwidth}
    \centering
    \includegraphics[width=\linewidth]{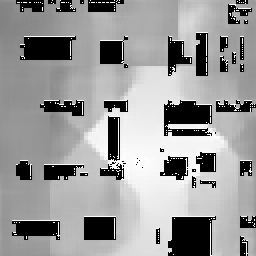}
    \caption{Predicted (UNet)}
    \label{Comparion:UNet}
  \end{subfigure}\hspace*{.025\textwidth}%
    \begin{subfigure}[t]{.205\textwidth}
    \includegraphics[width=\linewidth]{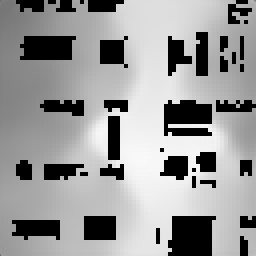}
    \caption{Predicted (RadioUNet)}
    \label{Comparion:RadioUNet}
  \end{subfigure}\hspace*{.025\textwidth}%
  \begin{subfigure}[t]{.205\textwidth}
    \centering
    \includegraphics[width=\linewidth]{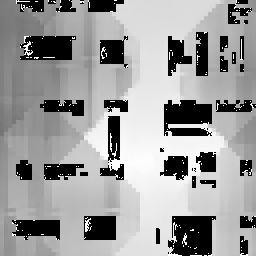}
    \caption{Predicted (PMNet)}
    \label{Comparion:PMNet}
  \end{subfigure}
  \caption{Comparison of the predicted pathloss map images of UNet, RadioUNet, and our proposed PMNet.}
  \label{fig:Comparison_Accuracy_Image}
  \vspace{-1.em}
\end{figure*}

\subsection{Comparison Study}
\textbf{Baselines}.\quad
This subsection compares the following three baselines, UNet, RadioUNet, and our proposed PMNet. 
All baselines here produce a single channel $256 \times 256$ image as the output, with the input of two channel $256 \times 256$ image where the first channel is the geographical map of RoI and the second channel is the TX location.
The detail of those baselines is summarized below. 

\begin{enumerate}
\item \textbf{UNet} \cite{ref:UNet} uses a simple U-shaped encoder-decoder networks model (called vanilla UNet). 
In terms of the architecture, $4$ encoder layers each consist of a Maxpool layer followed by two sets of \textmd{convolution}, \textmd{batch normalization}, and \textmd{ReLU} layers. The encoders are followed by $4$ decoder layers, each consisting of a transposed convolution layer followed by two set of convolution, batch normalization, and ReLU layers. All the encoder layers are concatenated with decoders at corresponding levels and vice versa.
\item \textbf{Radio UNet} \cite{ref:RadioUNet} is a NN also designed for pathloss map prediction. The NN architecture is based on the vanilla UNet but uses two UNet (\textit{e.g.}, double UNet). 
Here, each UNet consists of $8$ encoder layers, and each layer consists of convolution, ReLU, and Maxpool layer. Similarly, there are $8$ decoder layers, each consisting of transposed convolution followed by ReLU. 
The encoders and decoders are concatenated similarly to the vanilla UNet. 
Here, curriculum training is used; in the first training stage, the first UNet is trained for certain epochs, while the second UNet is frozen; in the second training stage, the second UNet is trained with the two-channel input features and with the output of the first UNet.  That is, the first UNet has two-channel inputs, and the second UNet has three-channel inputs. 

\item \textbf{PMNet} is our proposed NN architecture. 
This network applies several parallel atrous convolutions with different rates \cite{ref:Deeplabv3+} and the hourglass network \cite{ref:UNet} to make a network architecture more suited for our task, pathloss map prediction. 
Our encoder consists of $5$ ResNet-based layers. Each ResNet layer comprises several bottleneck layers consisting of convolution, batch normalization, and ReLU. 
The decoder consists of $7$ layers consisting of convolution, Maxpool, ReLU, transposed convolution, and ReLU. 
Skip connections are used between encoders and decoders.
\end{enumerate}

\textbf{Accuracy}.\quad
In Table \ref{Table_Comparison_Accuracy}, we provide the comparison results in terms of the accuracy of PMP for our baselines, including UNet, RadioUNet, and our proposed PMNet.
For both Exclusive and Random dataset, our proposed PMNet outperforms the other baselines, UNet and RadioUNet in terms of NMSE and NDL.
Here, recall that each pixel of RoI corresponds to the predicted power; thus, the values of NMSE and NDL indirectly show how accurately the method predicts the pathloss over each pixel. 

For PMP, the NN fulfills two main tasks. It first conducts semantic segmentation to distinguish if a certain pixel is an RoI area or not.\footnote{The semantic segmentation might be required for some, but not all, applications of NN-based PMP.} Then, for the pixel in the RoI area, this NN predicts a received power according to the TX location and the geographical map information around such a pixel. 
Namely, the NN is required to do semantic segmentation and exploit the contextual information (\textit{e.g.}, predict power over a pixel considering surrounding structure).
Fig. \ref{fig:Comparison_Accuracy_Image} shows the predicted images from each baseline, and each image shows how well each baseline conducts such tasks for PMP.
As shown in the figure, RadioUNet demonstrates more precise semantic segmentation performance than the UNet structure, but the received power prediction for RoI seems not precise (some parts are blurred).
On the other hand, although the proposed PMNet does not precisely segment the RoI as RadioUNet, overall, it predicts the received power more accurately than RadioUNet and the vanilla UNet.

\subsection{Impact of data set}

\textbf{Small dataset}.\quad
As discussed earlier, obtaining measurement data (by ray tracing simulation) is expensive in terms of time and computing resources; thus, PMP-oriented NN is required to work well even with a limited number of data.
To see if our proposed PMNet works well even on a small dataset (\textit{i.e.}, a small number of training and validation data), still surpassing the other two baselines, we configure a new data set with a size of 25 \% of \emph{Random} dataset (called as \emph{Random (Small)}), and train again on this small dataset. 

As shown in Table \ref{Table_Comparison_Small}, PMNet shows the best performance in both NMSE and NDL in \emph{Random (Small)} dataset case. 
Remarkably, the NMSE score of our proposed is close to the one of RadioUNet trained on a 4-fold larger dataset (see \emph{Random} in Table \ref{Table_Comparison_Accuracy}). 
Such results show that our proposed NN can efficiently learn necessary information even when given a small dataset. 
In other words, the PMNet has the potential to identify RoI and predict power from RoI even with limited data.

\begin{table}[!h]
\centering
\caption{Comparison results for \emph{Random (Small)} dataset.}
\resizebox{1.\columnwidth}{!}{\begin{minipage}[h]{0.85\columnwidth}
\centering
\label{Table_Comparison_Small}
\begin{tabularx}{1\linewidth}{l c c c c}
\toprule[1pt]
{Scheme} & Average $\mathrm{NMSE}$ & Average $\mathrm{NDL}$ \\
\cmidrule(lr){1-1} \cmidrule(lr){2-2} \cmidrule(lr){3-3}

UNet \cite{ref:UNet} & $0.0049796$  \; & $0.0456481$
\\
RadioUNet \cite{ref:RadioUNet} & $0.0019637$  \; & $0.0304550$
 \\
PMNet (Proposed) & $\textbf{0.0006752}$ \; & $\textbf{0.0177551}$
 \\

\bottomrule[1pt]
\end{tabularx}

  \vspace{-.5em}
\end{minipage}}
\end{table}

\textbf{Generalizability}.\quad
Table \ref{Table_Comparison_RadioMapseer} shows the comparison results, where the baselines are newly trained with \emph{RadioMapSeer} dataset, which is provided by \cite{ref:RadioUNet}.
All baselines are trained and validated with this different type of PMP dataset to see if our proposed PMNet is \emph{generalizable} for PMP tasks. Here, the generalization refers to performance over an unseen dataset \cite{deep_book}.
Table \ref{Table_Comparison_RadioMapseer} shows that our proposed outperforms even in the \emph{RadioMapSeer} dataset; that somewhat corroborates the generalizability of our proposed PMNet. 

\begin{table}[!h]
\centering
\caption{Comparison results for \emph{RadioMapSeer} dataset.}
\resizebox{1.\columnwidth}{!}{\begin{minipage}[h]{.85\columnwidth}
\centering
\label{Table_Comparison_RadioMapseer}
\begin{tabularx}{1\linewidth}{l c c c c}
\toprule[1pt]
{Scheme} & Average $\mathrm{NMSE}$ & Average $\mathrm{NDL}$ \\
\cmidrule(lr){1-1} \cmidrule(lr){2-2} \cmidrule(lr){3-3}

UNet \cite{ref:UNet} & $0.0031357$ \; & $0.0325763$
\\
RadioUNet \cite{ref:RadioUNet} & $0.0005134$ \; & $0.0103112$ 
 \\
PMNet (Proposed) & $\textbf{0.0004066}$ \; & $\textbf{0.0098763}$ 
 \\

\bottomrule[1pt]
\end{tabularx}

\end{minipage}}
\vspace{-1.em}
\end{table}

\subsection{Applications}

As seen in the results above, PMNet is an accurate large-scale channel prediction tool that captures propagation characteristics over surrounding environments. 
Not only does the PMNet allow for accurate channel prediction, but it also can be applied to several applications.
Here, we present one example of PMNet application, \emph{coverage prediction}.

\begin{figure}[h!]
  \centering
  \begin{subfigure}[t]{.22\textwidth}
    \centering
    \includegraphics[width=\linewidth]{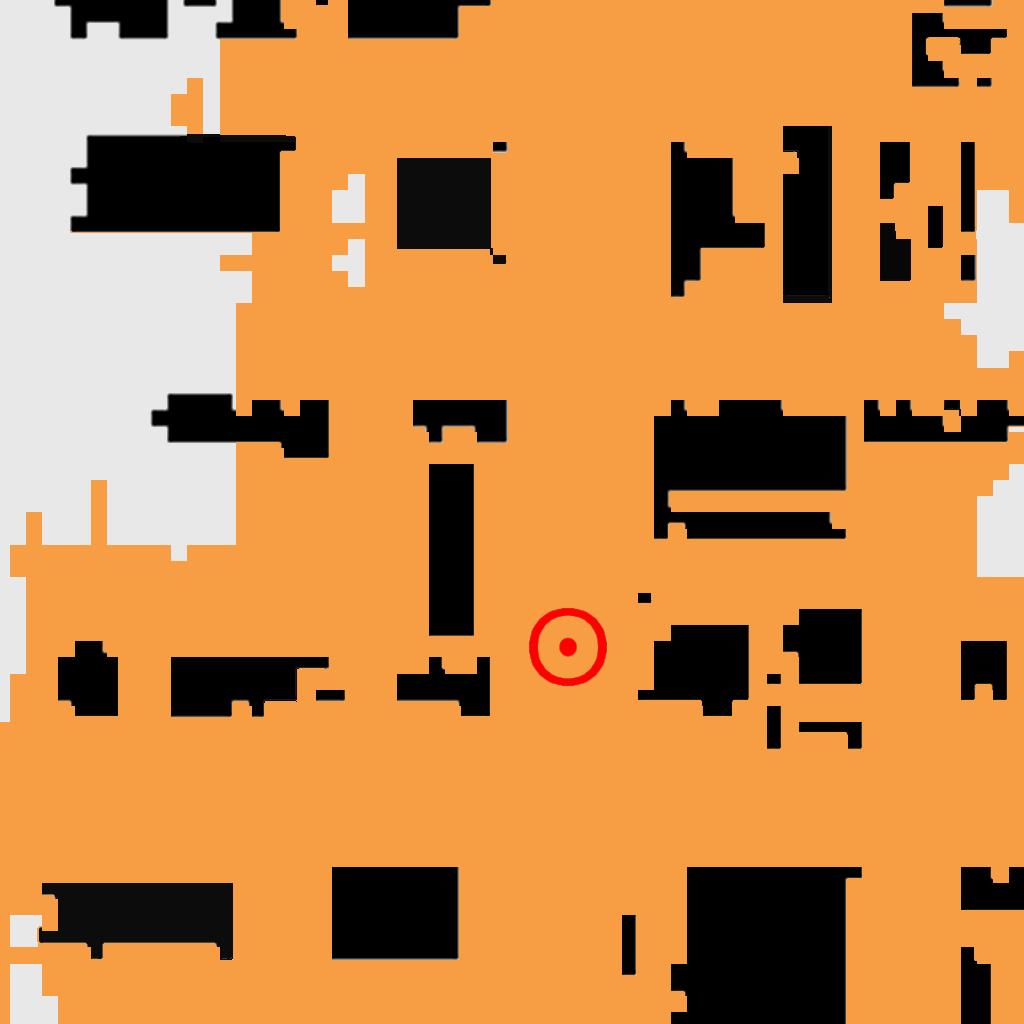}
    \caption{$\vb*P_{\mathrm{Thr}}= -90$ [dBm]}
  \end{subfigure}\hspace*{.025\textwidth}%
  \begin{subfigure}[t]{.22\textwidth}
    \centering
    \includegraphics[width=\linewidth]{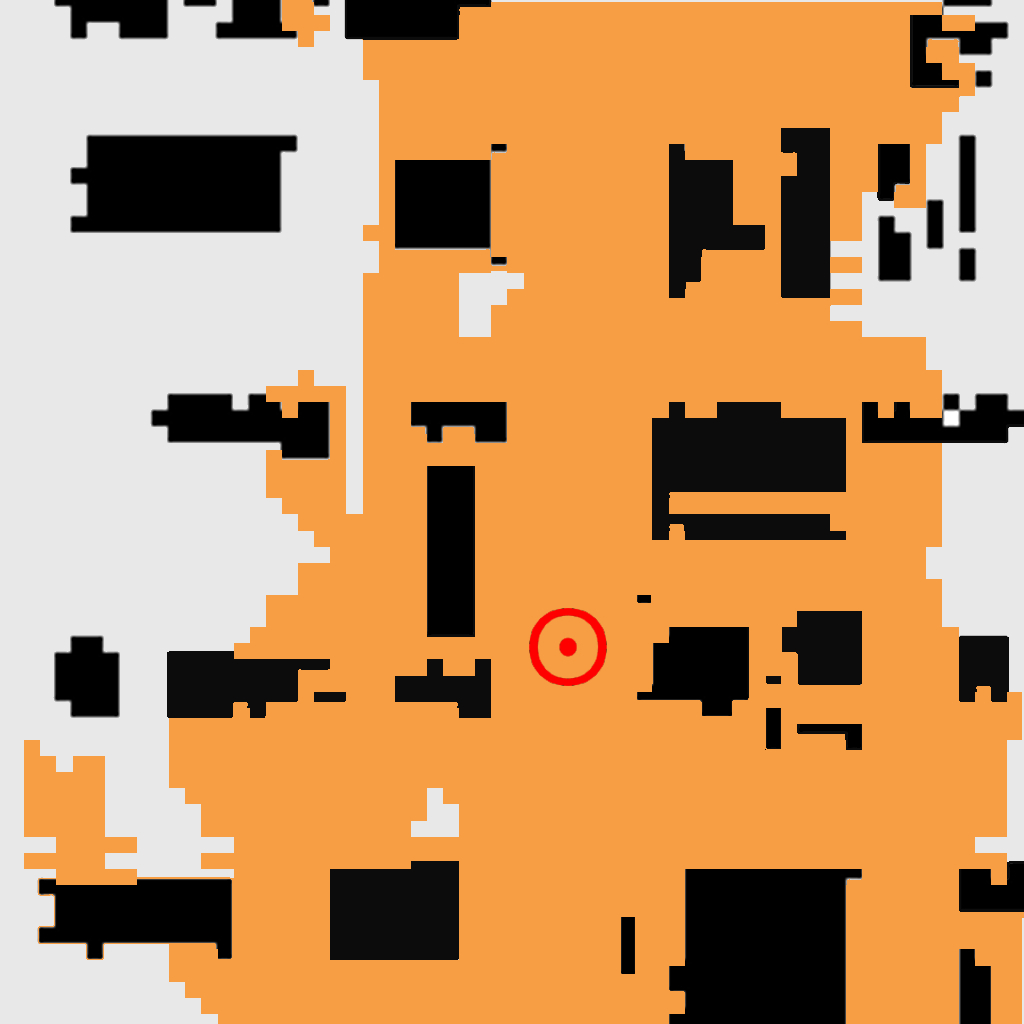}
    \caption{$\vb*P_{\mathrm{Thr}}= -80$ [dBm]}
  \end{subfigure} \vspace{.5em}\\
    \begin{subfigure}[t]{.22\textwidth}
    \centering
    \includegraphics[width=\linewidth]{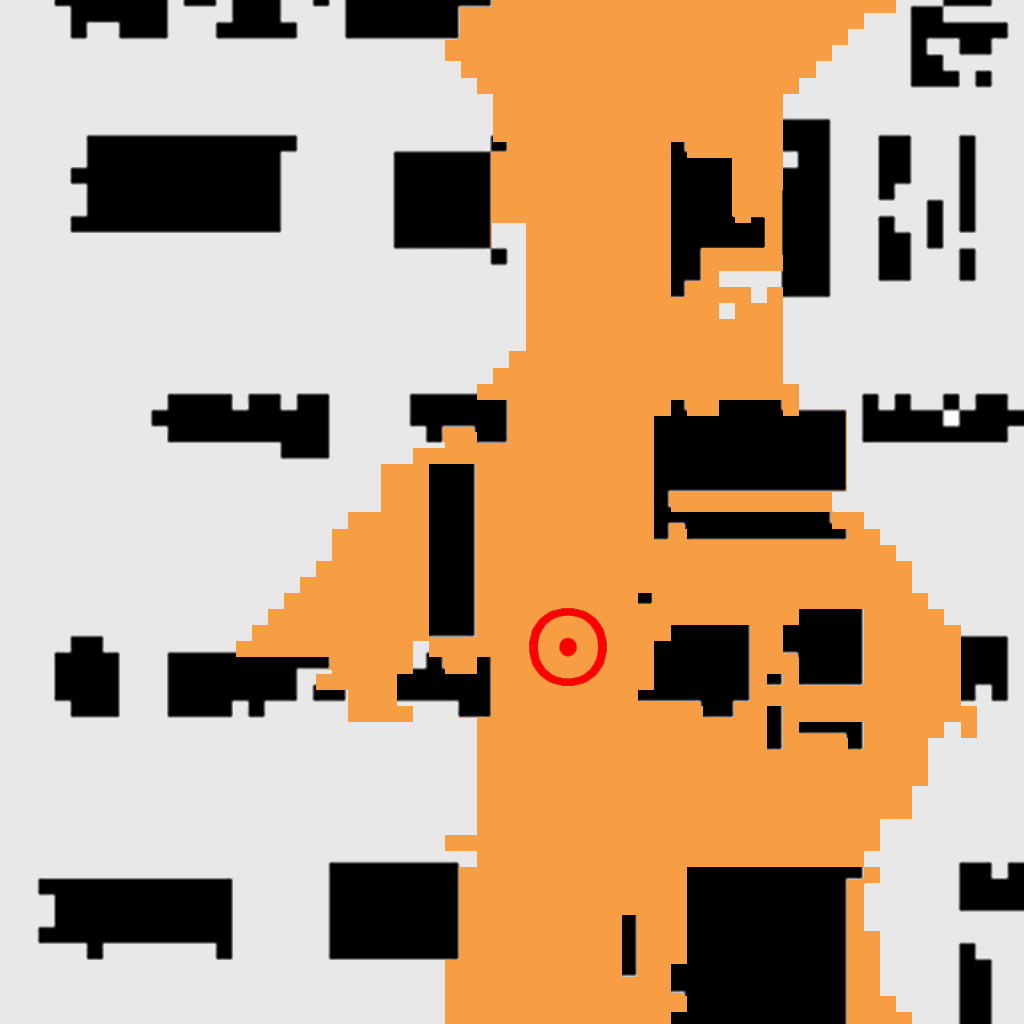}
    \caption{$\vb*P_{\mathrm{Thr}}= -70$ [dBm]}
  \end{subfigure}\hspace*{.025\textwidth}%
  \begin{subfigure}[t]{.22\textwidth}
    \centering
    \includegraphics[width=\linewidth]{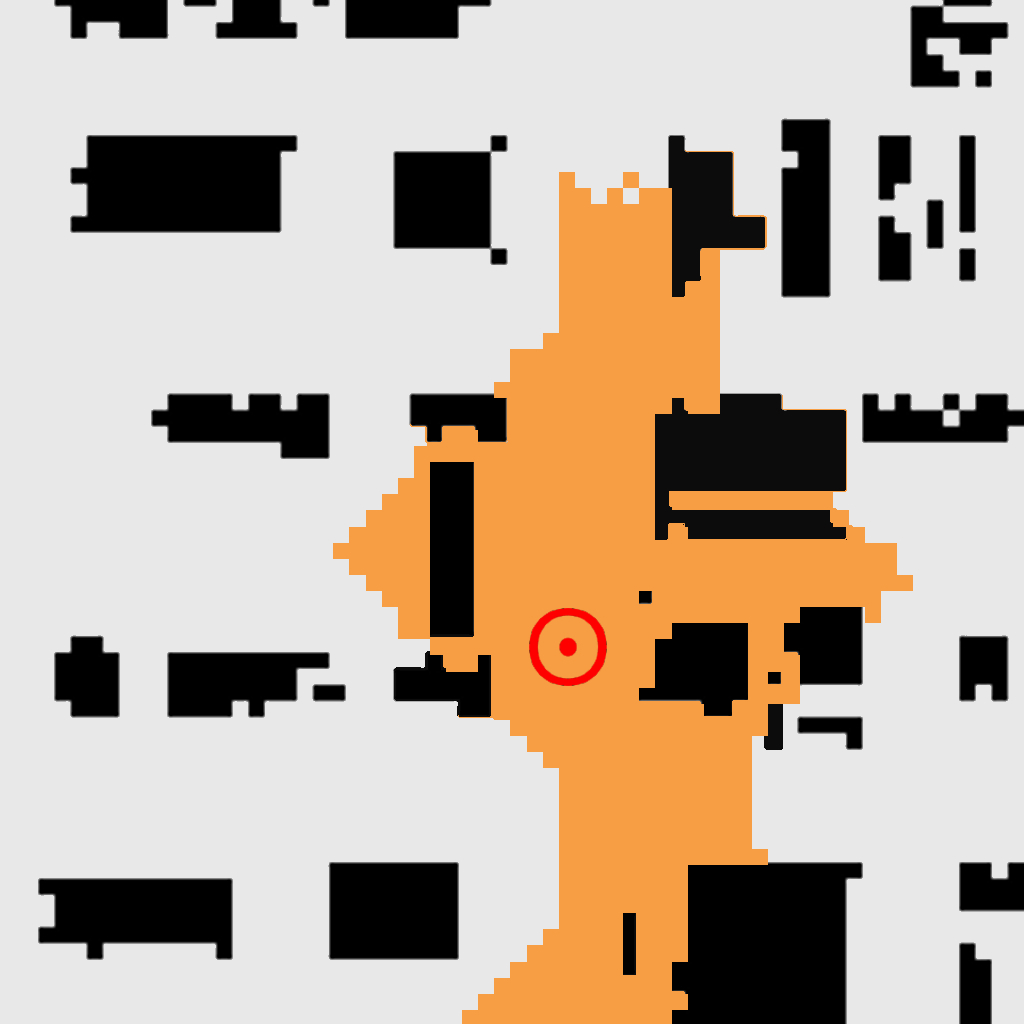}
    \caption{$\vb*P_{\mathrm{Thr}}= -60$ [dBm]}
  \end{subfigure}
  \caption{Coverage map with different required received power. The area with orange color is within the coverage, and the area with light grey color is outside of coverage. $\odot$ represents the location of TX.}
  \label{fig:Coverage Map}
  \vspace{-.5em}
\end{figure}

\textbf{Coverage map}.\quad
Fig. \ref{fig:Coverage Map} shows the coverage map drawn by the predicted received power over $i$-$j$ pixel, $\vb*P^{i, j}_{\mathrm{RX}}$, obtained from PMNet.
We simply define the coverage indicator of $i$-$j$ pixel $c^{i, j}$ as
\begin{align}
c^{i, j} = \begin{cases}
    1, & \vb*P^{i, j}_{\mathrm{RX}} \geq \vb*P_{\mathrm{Thr}}\\
    0, & \vb*P^{i, j}_{\mathrm{RX}} < \vb*P_{\mathrm{Thr}}
  \end{cases}, \label{Access}
\end{align}
where $c^{i, j}=1$ represents the $i$-$j$ pixel is within the coverage and vice versa, with $\vb*P_{\mathrm{Thr}}$ being the threshold for coverage.
One can observe the coverage hole in this figure with four different $\vb*P_{\mathrm{Thr}}$.
Intuitively, the performance of the coverage prediction depends on the accuracy of PMP.

\section{Conclusion} \label{conclusion}


We propose and evaluate a novel neural network (NN) architecture, PMNet, specifically designed for channel prediction tasks. By incorporating a unique combination of computer vision methods tailored to wireless propagation characteristics, PMNet outperforms existing architectures across various datasets. The method is robust, even with limited data, performing on par with a standard UNet trained on four times the amount of data. Additionally, PMNet demonstrates generalizability, maintaining accurate predictions when trained with datasets from different environments. We present a straightforward application of PMNet for coverage map prediction, while also recognizing its potential for other use cases.

{\bf Acknowledgements:} This work was funded by the NSF-AoF grant 2133655. The help of Dr. Zheda Li in the creation of the USC data set is gratefully acknowledged. 




\bibliographystyle{IEEEtran} 
\bibliography{Refs}

\begin{thebibliography}{10}
\providecommand{\url}[1]{#1}
\csname url@samestyle\endcsname
\providecommand{\newblock}{\relax}
\providecommand{\bibinfo}[2]{#2}
\providecommand{\BIBentrySTDinterwordspacing}{\spaceskip=0pt\relax}
\providecommand{\BIBentryALTinterwordstretchfactor}{4}
\providecommand{\BIBentryALTinterwordspacing}{\spaceskip=\fontdimen2\font plus
\BIBentryALTinterwordstretchfactor\fontdimen3\font minus
  \fontdimen4\font\relax}
\providecommand{\BIBforeignlanguage}[2]{{%
\expandafter\ifx\csname l@#1\endcsname\relax
\typeout{** WARNING: IEEEtran.bst: No hyphenation pattern has been}%
\typeout{** loaded for the language `#1'. Using the pattern for}%
\typeout{** the default language instead.}%
\else
\language=\csname l@#1\endcsname
\fi
#2}}
\providecommand{\BIBdecl}{\relax}
\BIBdecl

\bibitem{kim1999radio}
S.-C. Kim and et~al., ``Radio propagation measurements and prediction using
  three-dimensional ray tracing in urban environments at 908 {MHz} and 1.9
  {GHz},'' \emph{IEEE Trans. on Veh. Technol.}, vol.~48, no.~3, 1999.

\bibitem{WirelessInsite}
{REMCOM}, ``{Wireless Insite},''
  {\url{https://www.remcom.com/wireless-insite-em-propagation-software}}.

\bibitem{degli'Esposti2007}
V.~Degli-Esposti, F.~Fuschini, E.~M. Vitucci, and et~al., ``Measurement and
  modelling of scattering from buildings,'' \emph{IEEE Trans. on Antennas
  Propag.}, vol.~55, no.~1, pp. 143--153, 2007.

\bibitem{dominant}
R.~Wahl, G.~Wölfle, P.~Wildbolz, and F.~L, ``Dominant path prediction model
  for urban scenarios,'' in \emph{IST Mobile and Wireless Commun.}, 2005.

\bibitem{voronoi}
M.~Lee and D.~Han, ``Voronoi tessellation based interpolation method for
  {Wi-Fi} radio map construction,'' \emph{IEEE Commun. Lett.}, vol.~16, no.~3,
  pp. 404--407, 2012.

\bibitem{survey_radiomap}
\BIBentryALTinterwordspacing
D.~Romero and S.-J. Kim, ``Radio map estimation: A data-driven approach to
  spectrum cartography,'' 2022. [Online]. Available:
  \url{https://arxiv.org/abs/2202.03269}
\BIBentrySTDinterwordspacing

\bibitem{ref:RadioUNet}
R.~Levie, Ã.~Yapar, G.~Kutyniok, and G.~Caire, ``{RadioUNet}: Fast radio map
  estimation with convolutional neural networks,'' \emph{IEEE Trans. on
  Wireless Commun.}, vol.~20, pp. 4001--4015, 02 2021.

\bibitem{ref:Fadenet}
V.~V. Ratnam, H.~Chen, and et~al., ``Fadenet: Deep learning-based mm-wave
  large-scale channel fading prediction and its applications,'' \emph{IEEE
  Access}, vol.~9, pp. 3278--3290, 2021.

\bibitem{rmp_gan}
A.~Marey, M.~Bal, H.~F. Ates, and B.~K. Gunturk, ``Pl-gan: Path loss prediction
  using generative adversarial networks,'' \emph{IEEE Access}, vol.~10, p.
  90474–90480, 2022.

\bibitem{rmp_manhattan}
A.~Gupta, J.~Du, D.~Chizhik, R.~A. Valenzuela, and M.~Sellathurai, ``Machine
  learning-based urban canyon path loss prediction using 28 ghz manhattan
  measurements,'' \emph{IEEE Trans. on Antennas and Propagation}, vol.~70,
  no.~6, p. 4096–4111, Jun 2022.

\bibitem{image_texture}
S.~P. Sotiroudis, K.~Siakavara, G.~P. Koudouridis, P.~Sarigiannidis, and S.~K.
  Goudos, ``Enhancing machine learning models for path loss prediction using
  image texture techniques,'' \emph{IEEE Antennas and Wireless Propagation
  Letters}, vol.~20, no.~8, p. 1443–1447, Aug 2021.

\bibitem{deep_book}
I.~Goodfellow, Y.~Bengio, and A.~Courville, \emph{Deep Learning}.\hskip 1em
  plus 0.5em minus 0.4em\relax MIT Press, 2016,
  \url{http://www.deeplearningbook.org}.

\bibitem{alexnet}
A.~Krizhevsky, I.~Sutskever, and G.~E. Hinton, ``Imagenet classification with
  deep convolutional neural networks,'' in \emph{Advances in Neural Information
  Processing Systems (NIPS)}, vol.~25, 2012.

\bibitem{resnet}
S.~Xie, R.~Girshick, and et~al., ``Aggregated residual transformations for deep
  neural networks,'' in \emph{IEEE Conf. on Computer Vision and Pattern
  Recognition (CVPR)}, Honolulu, HI, Jul 2017, p. 5987–5995.

\bibitem{auteoncoders}
D.~E. Rumelhart, G.~E. Hinton, and R.~J. Williams, \emph{Learning Internal
  Representations by Error Propagation}.\hskip 1em plus 0.5em minus 0.4em\relax
  Cambridge, MA, USA: MIT Press, 1986, p. 318–362.

\bibitem{dim_redux}
G.~E. Hinton and R.~R. Salakhutdinov, ``Reducing the dimensionality of data
  with neural networks,'' \emph{Science}, vol. 313, no. 5786, pp. 504--507,
  2006.

\bibitem{vae}
D.~P. Kingma and M.~Welling, ``Auto-encoding variational bayes,''
  \emph{arXiv:1312.6114 [stat.ML]}, Aug. 2013.

\bibitem{ref:Deeplabv3+}
L.-C. Chen, Y.~Zhu, G.~Papandreou, F.~Schroff, and H.~Adam, ``Encoder-decoder
  with atrous separable convolution for semantic image segmentation,'' in
  \emph{Computer Vision – ECCV 2018}.\hskip 1em plus 0.5em minus 0.4em\relax
  Springer, 2018.

\bibitem{ref:UNet}
O.~Ronneberger, P.~Fischer, and T.~Brox, ``{U-Net}: Convolutional networks for
  biomedical image segmentation,'' in \emph{Medical Image Computing and
  Computer-Assisted Intervention}.\hskip 1em plus 0.5em minus 0.4em\relax
  Springer, 2015.

\bibitem{ref:gradientloss}
J.~Hu and et~al., ``Revisiting single image depth estimation: Toward higher
  resolution maps with accurate object boundaries,'' in \emph{IEEE Winter Conf.
  on Appl. of Computer Vision (WACV)}, 01 2019, pp. 1043--1051.

\bibitem{ref:Multiloss}
J.-H. Lee and C.-S. Kim, ``Multi-loss rebalancing algorithm for monocular depth
  estimation,'' in \emph{Computer Vision – ECCV 2020}.\hskip 1em plus 0.5em
  minus 0.4em\relax Springer, 2020.

\end{thebibliography}


\end{document}